\documentclass[11pt]{article}
\usepackage{fullpage}
\usepackage{multirow}
\usepackage{amsmath}
\usepackage{amssymb}
\usepackage{amsthm}
\usepackage{mathrsfs}
\usepackage{array}
\usepackage{subfigure}

\usepackage{graphicx}
\usepackage{latexsym,color,amssymb,times}

\newtheorem{theorem}{Theorem}[section]

\theoremstyle{definition}

\newtheorem{definition}[theorem]{Definition}

\newtheorem{claim}[theorem]{Claim}

\newcommand{\ignore}[1]{}
\newcommand{\remove}[1]{}

\begin{document}

\title{On the Interplay between Incentive Compatibility and Envy Freeness}
\author{
Edith Cohen\thanks{
	AT$\&$T Labs-Research, 180 Park Avenue, Florham Park, NJ.
}
\and
Michal Feldman\thanks{
	School of Business Administration and Center for the Study of Rationality,
The Hebrew University of Jerusalem.
	}
\and
Amos Fiat\thanks{
	The Blavatnik School of Computer Science, Tel Aviv University.
}
\and
Haim Kaplan \thanks{
	The Blavatnik School of Computer Science, Tel Aviv University.}
\and Svetlana Olonetsky\thanks{
The Blavatnik School of Computer Science, Tel Aviv University.
}
}
\date{}
\maketitle \thispagestyle{empty}

\begin{abstract}
We study mechanisms for an allocation of goods among agents, where agents have no incentive to lie about their true values (incentive compatible) and for which no agent will seek to exchange outcomes with another (envy-free). Mechanisms satisfying each requirement separately have been studied extensively, but there are few results on mechanisms achieving both. We are interested in those allocations for which there exist payments such that the resulting mechanism is simultaneously incentive compatible and envy-free.

Cyclic monotonicity is a characterization of incentive compatible allocations, local efficiency is a characterization for envy-free allocations.
We combine the above to give a characterization for allocations which are both incentive compatible and envy free. We show that even for allocations that allow payments leading to incentive compatible mechanisms, and other payments leading to envy free mechanisms, there may not exist any payments for which the mechanism is simultaneously incentive compatible and envy-free. The characterization that we give lets us compute the set of Pareto-optimal mechanisms that trade off envy freeness for incentive compatibility.
\end{abstract}

\newpage
\setcounter{page}{1}

\section{Introduction}

We consider allocation problems, where a set $U$ of objects should be allocated among $m$ agents, each having a valuation function $v_i$ assigning a value to every bundle. We wish to find a partition of the objects among the agents so as to achieve some goal. Typically, this goal is to maximize (or approximate) the social welfare, {\sl i.e.}, the sum of the agents' valuations for their bundles.
A mechanism $M=\langle a,p \rangle$ is a protocol that receives the set of the agents' valuations as input and returns a tuple consisting of an allocation $a$ and payments $p$ for the agents. The utility of an agent is the sum of valuations of the items minus any payments (quasi-linear utility).

Two natural desired properties of any mechanism are {\sl incentive compatibility} and {\sl envy-freeness}. A mechanism is incentive compatible if it is a dominant strategy for every agent to report her private information truthfully~\cite{Hurwicz}.
In 1987, Rochet \cite{Rochet} defined the notion of a cycle-monotonic allocation, and proved that this was a necessary and sufficient condition of an {\sl IC-implementable} allocation -- an allocation having associated payments that jointly form an incentive compatible mechanism.
A mechanism is envy-free if no agent wishes to switch her outcome with that of another~\cite{dubins,Foley,Svensson,Maskin87,Moulin04,Young95}.
The notion of locally efficient allocation has been defined by Haake {\sl et. al.} \cite{Haake2002} who showed that this was a necessary and sufficient condition of an {\sl EF-implementable} allocation -- allocation having associated payments that jointly form an envy-free mechanism.

While much work has been done on each of these properties independently, not much is yet understood about the interrelation between incentive compatibility and envy freeness The interaction between these two notions is the focus of our paper.

Our contribution is a characterization of allocations that are both incentive compatible and envy-free. We use this to derive various Pareto-optimal mechanisms that trade off envy freeness for incentive compatibility, we also use this to obtain negative results and to show separation results for different types of allocations and problems.

\subsection{Envy-Free and Incentive-Compatible Allocations}

\label{sec:ef_ic}
Motivated by the cyclic monotonicity characterization of \cite{Rochet} and the locally efficient characterization of \cite{Haake2002} we now consider two new categories of allocation functions:
\begin{description}

\item[$EF \cup IC$-implementable:]

An allocation function $a$ is called {\sl incentive compatible
    or envy-free implementable} ($EF \cup IC$-implementable) if there
  exists a payment function $p$ such that $M=\langle a,p \rangle$ is incentive
  compatible and a (possibly different) payment function $p'$ such that $M=\langle a,p' \rangle$ is
  envy-free.
\item[$EF \cap IC$-implementable:]

An allocation function $a$ is {\sl incentive
    compatible and envy-free implementable} ($EF \cap
  IC$-implementable) if there exists a payment function $p$ such that
  the mechanism $M=\langle a,p \rangle$ is incentive compatible and envy-free. Clearly, every function which is $EF \cap
  IC$-implementable is also $EF \cup IC$-implementable.
\end{description}
Many natural questions arise regarding the properties above: Are these classes identical? Empty? What interesting problems fall into each of these classes?
Figure \ref{fig:tab1} gives different payment functions for the social welfare maximizing allocation of one indivisible item. It is easy to verify the properties claimed for the various payment functions. The social welfare maximizing allocation is indeed  $EF \cap IC$, Clarke pivots payments give a mechanism that is both incentive compatible and envy free (the third entry).

In Section~\ref{sec:ga} we provide a characterization for  allocations that are
$EF\cap IC$-implementable. We then use the obtained characterization to derive some useful observations regarding the spectrum of ``best possible'' tradeoffs between (approximate) envy-freeness and (approximate) incentive compatibility.

Finally, in Section~\ref{sec:examples} we explore the relationships between the different classes. In particular, we show that not every $IC$-implementable allocation is also $EF$-implementable, and vice versa. Additionally, using the characterization provided in Section~\ref{sec:ga}, we show that not every $IC \cup EF$-implementable allocation is also $IC \cap EF$-implementable.

\begin{figure}
\begin{center}\begin{tabular}{|l|l|}
  \hline
  Payments & Properties \\
  \hline
  \hbox{\begin{tabular}{l}
    $ p_i = \min_{j\neq i} v_j$ \\
     $p_j = \min_{k\neq j} v_k - v_i, \mbox{\quad for all\ }j \neq i$
   \end{tabular}}
 & \hbox{\begin{tabular}{l}
    Incentive compatible, not envy-free \\
     (VCG, not Clarke pivot payments)   \end{tabular}} \\
     \hline
  \hbox{\begin{tabular}{l}
    $ p_i = 0$\\
     $p_j = -\max_{k\neq i} v_k$
   \end{tabular}} & \hbox{\begin{tabular}{l}
    Envy-free, not incentive compatible \\
     (not VCG)   \end{tabular}} \\
   \hline
   \hbox{\begin{tabular}{l}
    $ p_i = \max_{k\neq i} v_k$ \\
     $p_j = 0, \mbox{\quad for all\ } j \neq i$   \end{tabular}}& \hbox{\begin{tabular}{l}
    Envy-free and incentive compatible \\
     (VCG with Clarke pivot payments)   \end{tabular}} \\
  \hline
\end{tabular}\end{center}

\caption{Allocating a single item to the agent of highest valuation, (agent $i$), the utility of agent $i$ is $v_i - p_i$, the utility of
agents $j\neq i$ is $-p_j$.} \label{fig:tab1}
\end{figure}

\section{Preliminaries} \label{sec:def}

Let $U$ be a set of objects, and associated with agent $i$, $1\leq i
\leq m$, is a valuation function $v_i \in V_i$ that maps sets of
objects into
 $\Re$. Let $v=<v_1, v_2, \ldots, v_m>$ be a sequence of valuation
functions, and let $(v'_i,v^{-i})$ be the sequence of
valuation functions arrived by substituting $v_i$ by $v'_i$
($v'_i \in V_i$) , i.e.,
$$(v'_i,v^{-i}) = <v_1, \ldots, v_{i-1}, v'_i, v_{i+1},
\ldots, v_m>.$$

An allocation function\footnote{Here we deal with indivisible allocations, although our results also extend to divisible allocations with appropriate modifications.} $a$ maps a sequence of valuation functions
$v=<v_1, v_2, \ldots, v_m>$, into a partition of $U$ consisting of
$m$ parts, one for each agent. {\sl I.e.}, $$a(v) = <a_1(v), a_2(v),
\ldots, a_m(v)>,$$ where $\bigcup_i a_i(v) = U$ and $a_i(v) \cap a_j(v)
= \emptyset$ for $i \neq j$. Let $A$ denote the set of possible allocations. A payment function\footnote[6]{In this
paper we consider only
  deterministic mechanisms and can therefore omit the allocation as an
  argument to the payment function.} is a mapping from $v$ to $\Re^m$,
$p(v) = <p_1(v), p_2(v), \ldots, p_m(v)>$, $p_i(v)\in \Re$.
Payments are from the agent to the mechanism (if the payment is negative then this means that the transfer is from the mechanism to the agent).


A mechanism is a pair of functions, $M=\langle a,p \rangle$, where $a$ is an allocation function, and $p$ is a payment function. For
a sequence of valuation functions $v=\langle v_1, v_2, \ldots, v_m \rangle$, the utility to agent $i$ is defined as $v_i(a_i(v)) - p_i(v)$. Such a utility function is known as quasi-linear.

\paragraph{Vickrey-Clarke-Groves (VCG) mechanism:} A celebrated result in mechanism design is the family of \emph{Vickrey-Clarke-Groves} (VCG) mechanisms. A mechanism $M=\langle a,p \rangle$ is called a VCG mechanism if:
\begin{itemize}
\item[$\bullet$] $a(v) \in \mbox{\rm argmax}_{a \in A} \sum_{i=1}^m v_i(a_i(v))$, and
\item[$\bullet$] $p_i(v)= h_i(v^{-i}) - \sum_{j \neq i} v_j(a_j(v))$, where $h_i$ does not depend on $v_i$, $i=1,\ldots,m$.
\end{itemize}

For all $\{h_i\}_{i=1}^m$, the VCG mechanism is incentive compatible (See, {\sl e.g.}, \cite{Noam07}).
The \emph{Clarke pivot payment} for a VCG mechanism takes $$h_{i}(v^{-i})=\max_{a' \in A} \sum_{j \neq i} v_j(a').$$
We next define mechanisms that are IC, EF, or both IC and EF.

\begin{enumerate}
\item[$\bullet$]
A mechanism  is  {\em incentive compatible\/} if it is a dominant strategy for every agent to reveal her true valuation function to the mechanism. {\sl I.e.}, if for all $i$, $v$, and $v'_i$:
$$
v_i(a_i(v)) - p_i(v) \geq v_i(a_i(v'_i,v^{-i})) - p_i(v'_i,v^{-i});
$$
this is equivalent to:
\begin{equation}
\quad  p_i(v) \leq p_i(v',v^{-i}) + \Big(v_i(a_i(v)) - v_i(a_i(v'_i,v^{-i}))\Big).
\label{eq:iccond}
\end{equation}
\item[$\bullet$]
A mechanism  is  {\em envy-free\/} if no agent seeks to switch her allocation and payment with another. {\sl I.e.}, if for all $1\leq i,j\leq m$ and all $v$:
$$
v_i(a_i(v)) - p_i(v) \geq v_i(a_j(v)) - p_j(v);
$$
this is equivalent to:
\begin{equation}
\quad p_i(v) \leq p_j(v) + \Big(v_i(a_i(v)) - v_i(a_j(v))\Big).
\label{eq:efcond}
\end{equation}
\item[$\bullet$] A mechanism $(a,p)$ is {\sl incentive compatible and envy-free\/} if $(a,p)$ is both incentive compatible and envy-free.
\end{enumerate}

\section{Characterizing Allocations that are $EF\cap IC$-implementable}
\label{sec:ga}

Before presenting our characterization for allocations that are $EF\cap IC$-implementable, we present several known characterizations.

\begin{definition} (Locally Efficient(\cite{Haake2002})) An allocation function $a$ is said to be {\em locally efficient\/}  if for all $v$, and all permutations $\pi$ of $1, \ldots, m$, $$\sum_{j=1}^m v_i(a_i(v)) \geq \sum_{j=1}^m v_{i}(a_{\pi(i)}(v)).$$ \end{definition}

\begin{theorem} \label{thm:haake}
(\cite{Haake2002}) A necessary and sufficient condition for an allocation function $a$ to be $EF$-implementable is that $a$ is locally efficient.
\end{theorem}

\begin{definition} (Cycle monotonicity \cite{Rochet}) We require Rochet's notion of cyclic monotonicity : an
allocation function satisfies cycle monotonicity if for every player
$i$, every integer $K$, and every $v^1_i, v^2_i,\ldots, v^K_i \in
V_i$, we have
\begin{equation}
\sum_{k=1}^{K}\left[ v^k_i\left(a_i(v^k_i,v^{-i})\right) -
v^k_i\left(a_i(v^{k+1}_i,v^{-i})\right) \right] \geq 0
\end{equation}
where $v^{K+1}_i=v^1_i$. Note that the summand is the same as the expression in Equation (\ref{eq:iccond}).
\end{definition}
\begin{theorem} \label{thm:rochet} (\cite{Rochet}) A necessary and sufficient condition that an allocation function is $IC$-implementable is that it is cycle monotonic.
\end{theorem}

\subsection{The Graph $G_a$}
\label{sec:graph_ga}

For an allocation function $a$, let $G_a$ be a weighted digraph with vertices
$(i,v)$, where $1 \leq i \leq m$, and $v=<v_1, \ldots, v_m>$ is a sequence of
valuation functions.

The arcs of the graph are classified as either $EF$ arcs or $IC$ arcs as follows:

\begin{itemize}
\item {\bf $EF$ arcs:} For all $v$, all $1 \leq i,j\leq m$, there is an arc from $(j,v)$ to $(i,v)$ of weight
 $v_i(a_i(v))-v_i(a_j(v))$.
\item {\bf $IC$ arcs:} For all $1\leq i \leq m$, all $v^{-i}$, all $v_i$, and all $v_i'$, there is an arc from  $(i,(v_i',v^{-i}))$ to
$(i,(v_i, v^{-i}))$ of weight $v_i(a_i(v_i,v^{-i})) - v_i(a_i(v_i',v^{-i}))$.
\end{itemize}

 If we consider only $EF$ arcs, then $G_a$ consists of vertex disjoint complete digraphs, each of which corresponds to a different $v$.
 An allocation is $EF$-implementable if and only if there are no negative cycles of $EF$ arcs in $G_a$. This is equivalent to the locally efficient characterization of Theorem \ref{thm:haake}.

 If we consider only $IC$ arcs, then $G_a$ consists of vertex disjoint  complete digraphs, each of which corresponds to a different pair $i$ and $v^{-i}$. An allocation is $IC$-implementable if and only if there are no negative cycles of $IC$ arcs in $G_a$. This is  equivalent to the cycle monotone characterization of incentive compatible mechanisms of Theorem \ref{thm:rochet}.

This suggests the following characterization of allocations that are $EF \cap IC$-implementable:
 \begin{theorem} \label{thm:icandef}
Allocation function $a$ is incentive compatible and envy-free implementable ($EF \cap IC$-implementable)
if and only if there are no negative cycles in $G_a$.
  \end{theorem}
\begin{proof}
A mechanism $(a,p)$ is $EF$ and $IC$ if and only if there exists a payment function $$p(v)=<p_1(v), p_2(v), \ldots p_m(v)>$$ such that the following two conditions hold:
\begin{itemize}
\item For any $IC$ arc $e=((i,v'), (i,v))$, it holds that $v_i(a_i(v)) - p_i(v) \geq v_i(a_i(v')) - p_i(v')$;\\
which is equivalent to
$$
v_i(a_i(v)) - v_i(a_i(v')) \geq p_i(v) - p_i(v').
$$
\item For any $EF$ arc $e=((j,v), (i,v))$, it holds that $v_i(a_i(v)) - p_i(v) \geq v_i(a_j(v)) - p_j(v)$;\\
which is equivalent to
$$
v_i(a_i(v)) - v_i(a_j(v)) \geq p_i(v) - p_j(v).
$$
\end{itemize}
In $G_a$, $w(e)= v_i(a_i(v)) - v_i(a_i(v'))$ for an $IC$ arc $e=((i,v'), (i,v))$, therefore $w(e)\geq p_i(v)-p_j(v)$. Similarly, $w(e)= v_i(a_i(v)) - v_i(a_j(v))$ for an $EF$ arc $e=((j,v), (i,v))$, therefore $w(e) \geq p_i(v)-p_i(v')$. If we sum up these inequalities over the set of arcs forming a cycle (consisting of alternate $IC$ and $EF$ arcs), we get that the sum of arc weights must be non-negative as the right hand sides cancel out.

If $G_a$ does not contain a negative cycle, we can compute shortest paths from any arbitrary start vertex $(i,v)$, and
interpret the length of the shortest path from $(i,v)$ to a vertex $(j,v')$ as $p_j(v')$. Shortest paths obey the required condition.
\end{proof}

  It follows that an allocation is $EF \cup IC$-implementable (potentially different payments for incentive compatibility and for envy-freeness) if and only if
all negative cycles in $G_a$ include at least one $EF$ arc and at least
one $IC$ arc.
Figure~\ref{fig:8cycle} (in Appendix~\ref{sec:examples}) helps visualize $G_a$ by illustrating a
subgraph of a $G_a$ that contains a cycle of
$EF$ and $IC$ arcs that is not a union
of complete $IC$ and $EF$ cycles (this 8-node subgraph is the smallest subgraph
with this property.)

\section{Trading Envy for Truthfulness}

\subsection{Tradeoffs between approximate envy-freeness and approximate incentive compatibility}

\label{subsec:tradeoff}

  A mechanism $(a,p)$ has {\em $\Delta$-approximate envy-freeness} if
for all $v$, all $1
\leq i,j \leq m$, $v_i(a_i(v)) - p_i(v) \geq v_i(a_j(v)) - p_j(v) - \Delta$
(no agent envies another by more than $\Delta$).
Similarly, $(a,p)$ has {\em $\Delta$-approximate incentive compatibility}
if no agent has incentive to lie that is larger than $\Delta$,
{\sl i.e.}
for all $1\leq i\leq m$, for all $v$ and $v'_i$,
$v_i(a_i(v)) - p_i(v) \geq v_i(a_i(v'_i,v^{-i})) - p_j(v^{-i},v'_i) - \Delta$.

For a graph $G_a$ as defined in Section~\ref{sec:graph_ga} and a pair
$(c_{ef},c_{ic})$ of nonnegative values, we define
$G_a^{+(c_{ef},c_{ic})}$ to be such that $c_{ef}$ is added to the
weight of all $EF$ arcs and $c_{ic}$ is added to the weight of all
$IC$ arcs in $G_a$.

 A pair $(c_{ef},c_{ic})$ is {\em cycle-correcting} for $G_a$ if
$G_a^{+(c_{ef},c_{ic})}$ does not contain negative cycles.
 By computing shortest paths from an arbitrary node in
$G_a^{+(c_{ef},c_{ic})}$ we obtain payments (and a mechanism) with
$c_{ef}$-approximate envy-freeness and $c_{ic}$-approximate
incentive compatibility.
A cycle-correcting pair $(c_{ef},c_{ic})$ is {\em minimal} if there is no
cycle-correcting pair $(x,y)\not= (c_{ef},c_{ic})$ such that $x\leq c_{ef}$
and  $y\leq c_{ic}$.

 The set $T_a$ of minimal cycle-correcting pairs
defines a tradeoff between
approximate envy-freeness and approximate incentive compatibility of
the allocation function $a$.
If $a$ is $EF \cap IC$-implementable, then $T_a=\{(0,0)\}$.
If $a$ is $EF$-implementable (but not necessarily $IC$-implementable),
there is a point of the form $(0,c_{ic})\in T_a$. The
corresponding mechanism is envy-free and has the best possible
(that is, $c_{ic}$-approximate)  incentive compatibility
subject to envy-freeness.
Similarly, if $a$ is $IC$-implementable, there is a point $(c_{ef},0)\in T_a$
with a corresponding incentive compatible mechanism that has the best
possible (that is, $c_{ef}$-approximate)
envy freeness subject to incentive compatibility.
These ``best'' tradeoffs can be computed in time polynomial in the size
of $G_f$ (minimum ratio cycles~\cite{AMObook93}).

\subsection{Partition based on Trustworthiness}
\label{subsec:partition}
Assume that one can find either envy free or truthful prices, but not both. Then, one can find prices that enforce envy freeness for some of the agents, and truthfulness for the complement, and can choose how to classify the agents.

Consider the EF and IC arcs graph such that there are no negative cycles
of only IC or only EF arcs, but there might be a mixed cycle. In particular,
any negative cycle must include a vertex where the arc entering the vertex is
an EF arc and the arc exiting the vertex is an IC arc.

We do as follows:
We remove all IC arcs entering $(i,X)$ where $i$ is trusted, and remove all EF arcs entering $(i,X)$ where $i$ is untrusted.
Any negative cycle must include both trusted and untrusted agents (otherwise it is limited to
edges of one type). There are no EF edges from trusted to untrusted vertices, and IC edges only connect vertices associated with the
same agent. Thus, there is no cycle including both trusted and untrusted vertices.

It follows that no trusted agent will be envious, since envy is prevented by the incoming EF edges. Likewise, no untrusted agent has incentive to lie, this is guaranteed by the incoming IC edges\footnote{In fact, we could try to add more EF and IC edges, subject to there being no negative cycle, and get a wider set of envy/incentive-compatibility constraints. We could combine this idea with that of increasing the remaining edges by some $\Delta$, so as to get some guarantees on the envy of the untrusted agents and the incentive to lie of the trusted agents.}.

\section{Separation Examples}
\label{sec:examples}

\begin{figure}[h]
\begin{center}
\input{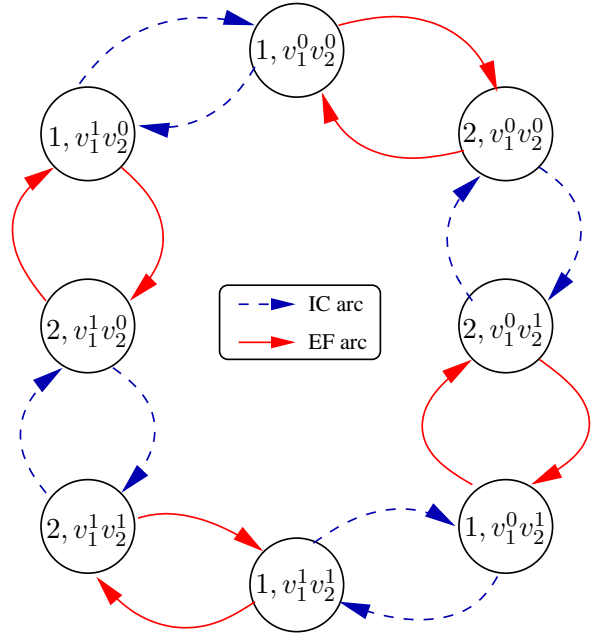}
 \caption{ A cycle in $G_a$ that includes two agents (1 and 2), and two valuation functions for each, $\{v_1^0, v_1^1\}$ for agent 1, $\{v_2^0,v_2^1\}$ for agent 2. The valuation functions for all other agents are fixed in this cycle.}\label{fig:8cycle}
\end{center}
\end{figure}

In this section we prove that not every allocation that is $IC$-implementable is also $EF$-implementable, and vice versa. Additionally, we show that not every allocation that is $IC \cup EF$-implementable is also $IC \cap EF$-implementable.

\begin{claim}
Not every allocation that is $IC$-implementable is also $EF$-implementable.
\end{claim}

\begin{proof}
Consider a single item auction and give it to agent $1$. If the mechanism pays nothing to any agent this is incentive compatible. However, there is no payment function that makes this envy-free, this allocation is not locally efficient (and thus not $EF$-implementable) unless agent $1$ has the highest valuation.
\end{proof}

\begin{claim}
Not every allocation that is $EF$-implementable is also $IC$-implementable.
\end{claim}

\begin{proof}
Consider a single divisible good. Let $a_i(v) = \alpha_i$ denote what fraction of the good is given to agent $i$, where  $\sum_{i=1}^m \alpha_i =1$. For the examples below, we take the valuation function to be proportional, {\sl i.e.}, for all $i$, $v_i(a_i(v)) = \alpha_i z_i$, $z_i$ is agent $i$'s value of the entire object.

Consider the following assignment function:
\begin{equation*}
\label{eq:abc}
a(v_1,v_2) =
\begin{cases}
(1/2,1/2) & z_1=z_2\\
(\frac{1}{4}-\frac{z_1}{2(z_1+z_2)}, \frac{3}{4}+\frac{z_1}{2(z_1+z_2)}) & z_1<z_2\\
(\frac{3}{4}+\frac{z_2}{2(z_1+z_2)},\frac{1}{4}-\frac{z_2}{2(z_1+z_2)}) & z_1>z_2\\
\end{cases}
\end{equation*}
This assignment is locally efficient, but the fraction assigned to an agent does not monotonically increase with the agent valuation. This contradicts cyclic monotonicity (on a cycle of length 2).
\end{proof}

\begin{claim}
Not every allocation that is $IC \cup EF$-implementable is also $IC \cap EF$-implementable.
\end{claim}

\begin{proof}
Consider a single divisible good and three agents $\{1,2,3\}$
where each of the agents, 1 and 2, can choose one of two specific valuation functions:
Agent $i=1,2,$ has valuation function $v_i\in\{v^{0}_i,v^{1}_i\}$, and agent 3 has only one
valuation function, $v_3$.

As before, let $z_i^j = v_i^j(1)$ (the valuation for the entire good).
Consider a setting where
$z^{0}_2 \geq z^{1}_2 > z^{0}_1 \geq z^{1}_1 \geq z_3\ .$
(For concreteness, take
$z_3=0$, $z^{0}_1=z^{1}_1=1$, $z^{0}_2=z^{1}_2=2$
for a divisible good and
$z_3=-3$, $z^{0}_1=z^{1}_1=-2$, $z^{0}_2=z^{1}_2=-1$
for a divisible task.)

We denote $a^{b_1 b_2}_i\equiv a_i(v^{b_1}_1,v^{b_2}_2,v_3)$
and consider the allocation
$a^{11}_1=a^{11}_2=0.4$, $a^{11}_3=0.2$ and
for $b_1 b_2 \in \{00,01,10\}$,
$a^{b_1 b_2}_1=a^{b_1 b_2}_2=0.5$, $a^{b_1 b_2}_3=0$.

We show that $a$ is $IC$-implementable (all $IC$ cycles in $G_a$ are nonnegative).
Any $IC$ cycle must involve one of the agents
$\{1,2\}$ (agent $3$ can not change valuations) and the
two valuations of this agent.  There are four $IC$
cycles (2-cycles) with weights, for $b_1 b_2 \in \{0,1\},$:
{\small
\begin{eqnarray*}
 (a^{0 b_2}_1-a^{1 b_2}_1)\cdot (z^{0}_1-z^{1}_1) &\geq& 0\\
 (a^{b_1 0}_2-a^{b_1 1}_2)\cdot (z^{0}_2-z^{1}_2) &\geq& 0
\end{eqnarray*}
}
(Using the property that for any agent $i\in\{1,2\}$, fixing the valuation of other agents, the fraction allocated is nondecreasing with valuation (weak monotonicity).)

One could argue directly that all EF cycles in $G_a$ are nonnegative and therefore $a$ is $EF$-implementable.
It is easier to see that $a$ is locally efficient. The two agents 1 and 2 always receive the same fraction and agent 3, whose valuation is smaller, receives a smaller fraction.

We now show that $G_a$ has a negative cycle, and therefore, using
Theorem~\ref{thm:icandef}, $a$ is not $EF \cap IC$-implementable.
Consider the 8-cycle $C_8$ (see Figure~\ref{fig:8cycle})
over players $1,2$ and valuations $(v^{b_1}_1,v^{b_2}_2,v_3)$ for
$b_1 b_2 \in \{00,01,10,11\}$.  The weight of $C_8$ is
{\small
\begin{eqnarray*}
 w(C_8)  & = & (a^{00}_1-a^{00}_2)\cdot z^{0}_1 +  \hspace{1cm}  \mbox{$\%$ $EF$ arc $(2,v_1^0v_2^0)\rightarrow (1,v_1^0v_2^0)$ }\\    
   && (a^{00}_2-a^{01}_2)\cdot z^{0}_2 + \hspace{1cm}  \mbox{$\%$ $IC$ arc $(2,v_1^0v_2^1)\rightarrow (2,v_1^0v_2^0)$}  \\ 
   && (a^{01}_2-a^{01}_1)\cdot z^{1}_2 + \hspace{1cm}  \mbox{$\%$ $EF$ arc $(1,v_1^0v_2^1)\rightarrow (2,v_1^0v_2^1)$ }  \\  
   && (a^{01}_1-a^{11}_1)\cdot z^{0}_1 + \hspace{1cm}  \mbox{$\%$ $IC$ arc $(1,v_1^1v_2^1)\rightarrow (1,v_1^0v_2^1)$ }  \\   
   && (a^{11}_1-a^{11}_2)\cdot z^{1}_1 +\hspace{1cm}  \mbox{$\%$ $EF$ arc $(2,v_1^1v_2^1)\rightarrow (1,v_1^1v_2^1)$ } \\  
   && (a^{11}_2-a^{10}_2)\cdot z^{1}_2 + \hspace{1cm}  \mbox{$\%$ $IC$ arc $(2,v_1^1v_2^1) \rightarrow (2,v_1^1v_2^0)$ } \\   
   && (a^{10}_2-a^{10}_1)\cdot z^{0}_2 + \hspace{1cm}  \mbox{$\%$ $EF$ arc $(1,v_1^1v_2^0)\rightarrow (2,v_1^1v_2^0)$ }  \\  
   && (a^{10}_1-a^{00}_1)\cdot z^{1}_1  \hspace{1.35cm}  \mbox{$\%$ $IC$ arc $(1,v_1^0v_2^0)\rightarrow (1,v_1^1v_2^0)$ }      
\end{eqnarray*}
}
This cycle has two arcs with nonzero weights (the 4th and the 6th arcs)
and has weight $w(C_8)=0.1(z^{0}_1-z^{1}_2)<0$.
\end{proof}

\bibliographystyle{plain}
\bibliography{envy}

\end{document}